\documentstyle[aps,pre,epsf,floats]{revtex}

\begin{document}

\title{
\draft Noise and Periodic Modulations in Neural Excitable Media
}
\author{
J. M. G. Vilar$^1$, R. V. Sol\'e$^{2,3}$, and J. M. Rub\'{\i}$^1$
}
\address{
$^1$Departament de F\'{\i}sica Fonamental, Facultat de
F\'{\i}sica, Universitat de Barcelona,
Diagonal 647, E-08028 Barcelona, Spain\\
$^2$Department of Physics, FEN, Universitat Polit\'ecnica de Catalunya,
Campus Nord, M\`odul B4,
E-08034 Barcelona, Spain\\
$^3$Santa Fe Institute, 1399 Hyde Park Road, Santa Fe,
New Mexico, 87501-8943\\
}
\maketitle
\widetext
\begin{abstract}
We have analyzed the interplay between noise and periodic modulations
in a mean field model of a neural excitable medium.
To this purpose, we have considered two types of modulations; namely,
variations of the resistance and oscillations of the threshold.
In both cases, stochastic resonance is present, irrespective
of if the system is monostable or bistable.
\end{abstract}


\section{Introduction}

When a nonlinear system is driven by
a periodic force in a noisy environment, its response
may be enhanced by the presence of noise.
This constructive role played by noise can be characterized
by the appearance of a maximum in the so called  signal-to-noise
ratio (SNR) at a nonzero noise level. In essence, the 
SNR is a quantity that reflects the quality of the
output signal, in such a way that for large values of this
quantity the output signal looks more ordered.
This phenomenon, named stochastic resonance (SR)
has been found in many situations pertaining
to different scientific areas \cite{Benzi,tri,Ma1,gam,Ma2,neu1,JSP,neu2,%
Moss,Wies,Wiese,array,thre1,grifo,phi4,prl3,prl2,prl1,thre2,BV}.

In regards to neural systems, many examples of  SR has been found
theoretically in single neurons \cite{sn1,sn2} and neural
networks \cite{nn1,nn2,nn3}, and experimentally in single neurons \cite{esn1,esn2}.
In contrast, an aspect which has only
been considered recently is the appearance of SR
in neural excitable media \cite{nem}. In a  set of experiments with mammalian
brain slices corresponding to the hippocampal center \cite{carn}
it was
demonstrated that an electric field can either suppress or enhance coherent
activity in real networks.
It was later shown, by using a time-varying electric
field, that as the magnitude of the stochastic component of the field was
increased, SR was observed in the response of the neural network to a
weak periodic signal \cite{nem}. These and other recent results \cite{noineum}
clearly shows that
neural noise could play a relevant role in the information processing of the
brain.

 In this context, both key ingredients for SR,
i.e. a well-defined coherent
and time-periodic modulation and intrinsic noise, are present at several
scales in neural tissues.
We can consider a coarse-grained characterization of brain dynamics by
means of the analysis of the electroencephalogram, which is an
averaged measure of the spatiotemporal activity of millions of neurons.
These neurons, in turn, are part of a network receiving inputs from various
parts of the nervous system called nuclei.
Among these, the thalamus plays a very
important role in controlling the behavioral states
of the brain \cite{kan}. In fact,
the thalamus is known to display autonomous
oscillations \cite{kan}, i.e. it works as
some kind of pacemaker to the brain cortex. 
The thalamus is massively connected with the
cortex and produces autonomous periodic oscillations, even
when disconnected from the cortex. 
Hence, the cortex can be, to some extent, viewed as a highly connected network
which is periodically stimulated by the thalamic pacemaker. Several
experimental results give support to the idea that the behavioral states of
the brain (alpha rhythm, slow delta waves, sleep, and REM)
are somehow related
to an oscillatory input into the cortical tissue\cite{db}.
These facts indicate that
a realistic model of cortical dynamics should consider the effects of
the thalamo-cortical pacemaker, which
can be simulated in different ways,
but it can always be considered as a periodic external signal.

Another aspect to be emphasized is the fact that
single neurons of the cortex exhibit some degree of variability,
i.e. the response to a stimulus depends on the particular trial,
which is observed in single-neuron experiments
and also in measurements of single-neurons inside the brain. 
Such a variability comes from both complex deterministic dynamics and the
noise implicit in the random nature of the incoming signals.
Additionally, single neurons manifest intrinsic noise.
In this sense,
there have been numerous
experimental studies about the stochastic activity of nerve cells. Noise has
been observed in nerve-cell preparations and involves both synaptic noise,
which is due to randomly occurring synaptic potentials
\cite{tuc},
and
membrane noise. In the last case,
small fluctuations in the
electric potential across the nerve-cell membrane are observed,
even when apparently
steady conditions prevail.
These fluctuations are
linked with conductance changes induced by random closing and opening of
ion channels.

The aim of this paper is to study a simple model which is able
to capture the main traits of the actual neural media, as
concerns the interplay between noise and periodic modulation
in neural dynamics.
To this purpose, we present a detailed analysis
of a standard mean field model of neural excitable medium
which was introduced in a
preliminary form in Ref. \cite{prl1}.

The paper has been organized as follows: In section \ref{sec2} we present
the mean field model. In sections  \ref{sec3} and  \ref{sec4}, we analyze
the appearance of SR in two different situations concerning
oscillations of the resistance of the neural tissue and variation
of the threshold.
Finally, in section \ref{sec5} we summarize our main results and discuss
possible implications in neural systems.

\section{Mean Field Model}\label{sec2}

Let us consider a slab of neural tissue comprising a very large number
of closely packed and coupled nerve cells, where connections are only
excitatory.
Different parts of the neocortex can fit
this description with more or less success. In general,
most of the real networks formed in the brain cortex are
constituted by both excitatory and inhibitory neurons.
The frequency and relevance of each type of cell, however,
vary depending on the place considered.
In this sense, the overall activity of
some brain regions, like the thalamus, is largely
dominated by the excitatory component \cite{she},
usually identified, at the microscopic level, with the
piramidal cells. This excitatory behavior is
particularly relevant in the hippocampus,
when stimulation of a single excitatory
neuron can generate a burst of synchronous
activity \cite{kno}.
These results makes 
the consideration of  the behavior of cortical nets in terms
of purely excitable dynamics reasonable, as a first approximation.
In addition, the lack of a microscopic
structure is not
an important problem when neurons and connections are not
explicitly considered as discrete entities,
as done at our level of description.
In this regard, previous studies reveal that the mean-field approach
is able to reproduce  the detailed macroscopic description
\cite{sch}.

The situation  we consider involves one of the simplest
models of aggregates of nerve
cells. At a given spatial point $\vec r$, the quantity of interest is the
mean local potential $V({\vec r})$ which is the result of a local
integration of incoming signals, i. e.
\begin{equation}
V({\vec r},t)= \int_{\Gamma} v({\vec r}-{\vec {r^\prime}},t)
\rho({\vec {r^\prime}}) 
d{\vec {r^\prime}} \;\;,
\end{equation}
over a given neighborhood $\Gamma$. Here $\rho({\vec r})$ is the local
packing density of neurons
and $v({\vec r},t)$ the transmenbrane potential. 

The time evolution of $V({\vec r},t)$ can be obtained from the following
integrodifferential equation
\begin{equation}
C{\partial V({\vec r},t) \over \partial t} 
= - { V({\vec r},t) \over R} +
\beta ({\vec r}) \Phi \left(
 V({\vec r},t) 
\right) +I({\vec r},t) \;\;.
\end{equation}
Here C is the  capacitance, R the resistance,
$I({\vec r},t)$ a  stimulus and $\beta({\vec r})$
the mean number of synaptic connections. The functional
form of $\Phi$ is defined
through the sigmoidal relation:
\begin{equation}
\label{sigmo}
\Phi(V) = \Phi_0 \left[ 1 + e^{-\nu (V-\theta)} \right]^{-1} \;\;,
\end{equation}
where $\Phi_0$ and $\nu$ are given constants and $\theta$ is a threshold.
Many possible 
functional forms for $\beta({\vec r})$ can be considered,
as for example an exponential
decay $\beta ({\vec r}) = b \exp(- |{\vec r}|/\gamma)$ which leads to
the equation:
\begin{eqnarray}
\nonumber
C{\partial V({\vec r},t) \over \partial t}  & = & 
- { V({\vec r},t) \over R} \\
\nonumber
 &+& b \Phi_0 \int e^{-|{\vec r}-{\vec {r^\prime}}|/ \gamma} 
\left[ 1 + e^{-\nu (V({\vec {r^\prime}},t)-\theta)} \right]^{-1}
d{\vec {r^\prime}} \\
&+& I({\vec r},t) \;\;.
\end{eqnarray} 

We will consider the case of all-to-all connectivity, i. e. 
$\gamma \gg L$, where $L$ is the characteristic length scale.
Under this assumption,
the model of neural excitable medium constitutes a mean field approximation
and is formulated by the equation\cite{Cow1}
\begin{equation}
C{dV \over dt}=-R^{-1}V+\tilde \varepsilon(1+e^{-\nu(V-\theta)})^{-1} +I
 \;\; ,
\end{equation}
accounting for the dynamics of the spatial average $V$ of the transmembrane
potential.
Here $\tilde\varepsilon$ is a constant that arises from that approximation.
This model is referred to as the Cowan-Ermentrout model (CEM).
Its dynamics can be described through a
potential function $U(V)$,
\begin{equation}
{d V \over d t} = -{\partial U \over \partial V} + {I \over C} \;\;. 
\end{equation}
In our case, 
\begin{equation}
\label{pf1}
U(V)={1 \over 2\nu C}\left[\nu R^{-1}V^2 - 2\tilde \varepsilon
\ln\left(1+e^{\nu\left(V-\theta\right)}\right)\right] \;\; .
\end{equation}
A remarkable fact is that
for some values of the parameters, CEM exhibits bistability.
Therefore, small
changes in the parameter values can lead to sudden
shifts from one stable branch to the
other \cite{Cow1}.

To render our analysis of the mean field model complete,
we need to specify the nature of the noise.
It is worth pointing out that its origin, its characterization, and
its effects in actual neural media inside
the brain is far from being clear.
Here, in order to account for the noise effects,
we will consider a simplified situation
that can be described by a fluctuating current applied to the net.
Moreover, we assume that this current
may be approximated by a Gaussian white noise [$\left<I(t)\right>=0$ and
$\left<I(t) I(t+ \tau) \right> = 2\sigma \delta(\tau)$].

Another remarkable fact is that
periodic modulations may affect the system in different ways,
for instance, by periodically changing the value of one of
the parameters.
Thus, small changes in
the permeability of a suitable ion gives rise to
variations of the membrane resistance $R$. Electric fields
may cause shifts in the effective threshold $\theta$ for an action
potential initiation. The parameter $\theta$
could also change when stimulus are sent to the
medium from other regions, e.g. when two networks interact.
Here we will explicitly consider these two cases, i.e.
oscillations of the resistance and variations of
the threshold, although other possibilities could also
be analyzed.

\section{Oscillations of the resistance.}\label{sec3} 

As a  source of periodic modulation,
we will first focus on the oscillations of the
membrane resistance.
In order to proceed with our analysis, we will consider
the Fokker-Planck equation giving the probability density of
the spatial averaged transmenbrane potential.
For the sake of simplicity, we introduce the new set of
variables: $x\equiv V$, $x_0\equiv \theta$,
$k(t)\equiv R(t)^{-1}C^{-1}$, $D\equiv\sigma/C$,
and $\varepsilon \equiv \tilde \varepsilon / C$.
In these variables, the Fokker-Planck equation reads
\begin{equation} \label{FP1}
{\partial P \over \partial t} = -{\partial \over \partial x}
\left[(F_0+\varepsilon F_1)P\right]
+ D{\partial^2 P \over \partial x ^2} \;\; ,
\end{equation}
where $F_0=-k(t)x$, with $k(t)=\kappa[1+\alpha\cos(\omega_0 t)]$,
is the linear force and
$F_1 = [1+e^{-\nu (x-x_0)}]^{-1}$ accounts for
the nonlinear contribution. $\kappa$, $\alpha$, and $\omega_0$
are constant parameters.

The probability density can be expanded
in powers of the strength of
the nonlinear term; namely,
$P=P_0+\varepsilon P_1 + \varepsilon^2 P_2 + \dots$. 
Here $P_0$ corresponds to the linearized
equation 
whereas the remaining terms account for  corrections due 
to the nonlinearity. Substituting this expansion
in Eq. (\ref{FP1}) we obtain the evolution equations
for the different contributions:
\begin{eqnarray}
\label{P1}
{\partial P_0 \over \partial t} &=& -{\partial \over \partial x}
(F_0 P_0)
+ D{\partial^2 P_0 \over \partial x ^2}\;\; , \\
\label{P2}
{\partial P_1 \over \partial t} &=& -{\partial \over \partial x}
\left(F_0P_1+F_1P_0\right)
+ D{\partial^2 P_1 \over \partial x ^2}  \;\; , \\
\label{P3}
{\partial P_2 \over \partial t} &=& -{\partial \over \partial x}
\left(F_0P_2+F_1P_1\right)
+ D{\partial^2 P_2\over \partial x ^2}  \;\; .
\end{eqnarray}
To proceed further, we will define the quantities
\begin{equation}
\left<x_t\right>_i=\int_{-\infty}^\infty xP_i(x,t)dx \;\; ,
\end{equation}
with $i=1,2,3$, and
\begin{equation}
\left<x_{t+\tau}x_{t}\right>_{i,j}
=\int_{-\infty}^\infty xdx 
\int_{-\infty}^\infty yP_i(y,t+\tau \vert x,t)P_j(x,t)dy \;\;,
\end{equation}
with $i,j=1,2,3$.
From Eqs. (\ref{P1})-(\ref{P3})
one can easily see that $\left<x_t\right>_0=0$,
whereas  $\left<x_t\right>_1\neq 0$ and $\left<x_t\right>_2\neq 0$.
Moreover, the correlation function can be expanded in
the form
\begin{eqnarray}
\nonumber
\left<x_{t+\tau}x_{t}\right>
&=&\int_{-\infty}^\infty xdx 
\int_{-\infty}^\infty yP(y,t+\tau \vert x,t)P(x,t)dy \\
\nonumber
&=&\left<x_{t+\tau}x_{t}\right>_{0,0}\\
\nonumber
&+&\varepsilon\left<x_{t+\tau}x_{t}\right>_{1,0}
+\varepsilon\left<x_{t+\tau}x_{t}\right>_{0,1} \\
\nonumber
&+&\varepsilon^2\left<x_{t+\tau}x_{t}\right>_{1,1}\\
&+&\varepsilon^2\left<x_{t+\tau}x_{t}\right>_{2,0}
+\varepsilon^2\left<x_{t+\tau}x_{t}\right>_{0,2} \;\; .
\end{eqnarray}

To analyze the interplay between noise and input signal,
we will consider the SNR, defined as usual by
\begin{equation} \label{SNR}
\mbox{SNR}={S(\omega_0) \over N(\omega_0)}
\;\; ,
\end{equation}
where the output noise is in first approximation given by
\begin{equation}\label{noise}
N(\omega) = {2\pi \over \omega_0}\int^{\omega_0/2\pi}_0
\int^\infty_{-\infty}\left<x_{t+\tau}x_{t}\right>_{0,0}
cos(\omega\tau)d\tau dt
\;\; ,
\end{equation}
and the output signal $S(\omega_0)$ comes from
\begin{eqnarray}
\nonumber
S(\omega_0)\left[\delta(\omega_0-\omega)+\delta(\omega_0+\omega)\right] = \\
\label{signal}
={2\pi \over \omega_0}\int^{\omega_0/2\pi}_0  
\int^\infty_{-\infty}\varepsilon^2
\left<x_{t+\tau}\right>_1\left<x_{t}\right>_1
cos(\omega\tau)d\tau dt
\;\; .
\end{eqnarray}
Since for large $\tau$ the quantities $x_{t+\tau}$ and $x_t$ become
uncorrelated,
in the previous equation we have replaced $\left<x_{t+\tau}x_{t}\right>_{i,j}$
by $\left<x_{t+\tau}\right>_i\left<x_{t}\right>_j$ due to the fact that
to compute the signal it is sufficient to know the behavior of
the correlation function for larger times.
Therefore, to obtain the SNR it is sufficient to consider only
$P_0(y,t+\tau \vert x,t)$, $P_0(x,t)$ and $P_1(x,t)$.

With the purpose of obtaining the equation for the probability
density, we will assume that the resistance varies
slowly and that the amplitude
of the oscillations is small.
By using these approximations
and by taking into account the fact that when the 
contribution proportional to $\alpha$ can be neglected
the linear system constitutes an Ornstein-Uhlenbeck process,
the noise term can be readily computed from
\begin{equation}
\left<x_{t+\tau}x_{t}\right>_{0,0}={ D \over \kappa}e^{-\kappa\tau}
\;\; .
\end{equation}
Therefore the spectral density of the output noise is 
\begin{equation}
N(\omega)={ D \over 2\kappa ^2}
\;\; ,
\end{equation}
provided that $\omega \ll \kappa $.

Moreover, we will also assume that the sigmoidal function [Eq. (\ref{sigmo})],
giving the
mean firing rate, may be approximated by a step function.
This fact occurs when the  gain of the neuron $\nu$
is sufficiently large.
In such a case, the potential function is given by
\begin{equation}
U(V)={1 \over 2C}\left[R^{-1}V^2
-2\varepsilon (x-x_0)\Theta(x-x_0)\right] \;\; .
\end{equation}
Under these circumstances,
\begin{equation}
\label{pp1}
P={1 \over Z}e^{-k(t)x^2/2D}e^{\varepsilon (x-x_0)\Theta(x-x_0)/D}
\;\; ,
\end{equation}
where $Z$ is the normalization factor.
Up to order $\varepsilon$ Eq. (\ref{pp1}) yields
\begin{equation}
P = P_0
+ \varepsilon P_1
\;\; ,
\end{equation}
where
\begin{equation}
P_0=\sqrt{k(t) \over 2\pi D}e^{-k(t)x^2 /2 D}
\;\; ,
\end{equation}
and
\begin{equation}\label{P1a}
P_1={1 \over D }P_0\left[(x-x_0)\Theta(x-x_0)
-\int^\infty_{x_0}(x-x_0)P_0dx\right] \;\; .
\end{equation}
Notice that
by using the fact that $\alpha$ is small $P_0$ can
be approximated by
\begin{equation} \label{P0a}
P_0 = \sqrt{\kappa  \over 2\pi D}
e^{-\kappa x^2 /2 D}\left[1+{1 \over 2}(1-x^2\kappa  /D)
\alpha\cos(\omega_0 t)\right]
\;\; .
\end{equation}

In order to obtain the output signal, we will take into account the
expression
\begin{equation}
\label{sus}
\left<x_t\right>_1={\cal B}+{\cal A}\alpha\cos(\omega_0 t) 
\;\; ,
\end{equation}
which holds when $\alpha$ and $\omega_0$ are sufficiently small.
Here ${\cal B}$ and  the ${\cal A}$ do not depend on time.
By using this expression in
Eq. (\ref{signal}) we obtain the signal as a
function of the susceptibility ${\cal A}$,
\begin{equation}
S(\omega_0)={\pi \over 2}(\alpha\epsilon{\cal A})^2
\;\; .
\end{equation}
This quantity can be computed from
Eqs. (\ref{P0a}) and (\ref{P1a}), and one obtains: 
\begin{equation}
{\cal A} = \int^\infty_{x_0}{1 \over 2D}\sqrt{\kappa  \over 2\pi D}
e^{-\kappa x^2 /2 D}x(x-x_0)(1-x^2\kappa  /D)dx
\;\; .
\end{equation}

The SNR then reads
\begin{equation}
\mbox{SNR}= \pi(\alpha\epsilon{\cal A})^2 {\kappa ^2\over D} 
\;\; .
\end{equation}
The previous integral cannot be performed explicitly, however,
its behavior for high and low noise levels can be obtained
easily.

The high noise level case
can be performed by replacing the lower limit of
the integral $x_0$ by $0$, provided that $x_0^2 \ll D/\kappa $.
We then obtain
\begin{equation}
{\cal A} ={1 \over 2D}\sqrt{\kappa  \over 2\pi D} \int^\infty_0
e^{-\kappa x^2 /2 D}\left(x^2-{\kappa  \over D}x^4\right)dx=-{1 \over \kappa }
\;\; .
\end{equation}
Therefore
\begin{equation}
S(\omega_0)={\pi \over 2}({\alpha\epsilon \over \kappa })^2
\;\; ,
\end{equation}
and
\begin{equation}
\mbox{SNR}= \pi(\alpha\epsilon)^2 {1 \over D} 
\;\; .
\end{equation}

In the same way, for low noise level we can perform
an asymptotic expansion by using the formula
\begin{equation}
\int_x^\infty u^n e^{-au^2}du=
{1 \over 2}x^{n+1} {e^{-ax^2} \over  ax^2}\left[1+{n-1 \over 2 ax^2}
+\mbox{O}\left({1 \over a^2}\right)\right] \;\; .
\end{equation}
In this case the susceptibility is given by
\begin{equation}
{\cal A} = -x_0\sqrt{1 \over 2\pi \kappa  D}
e^{-\kappa x_0^2/ 2D}
\;\; .
\end{equation}
Therefore, the signal and the SNR are 
\begin{equation}
S(\omega_0)
={1 \over 4\kappa D}(x_0\alpha\epsilon)^2e^{-\kappa x_0^2/ D}
\;\; ,
\end{equation}
and
\begin{equation}
\mbox{SNR}
={1 \over 2}\left(x_0{\alpha\epsilon \over D}\right)^2
\kappa e^{-\kappa x_0^2/ D}
\;\; .
\end{equation}
At low noise level the SNR increases whereas it decreases for large values
of the noise, therefore  the SNR has a maximum which indicates the 
presence of SR.

\begin{figure}[th]
\centerline{
\epsfxsize=8.0cm 
\epsffile{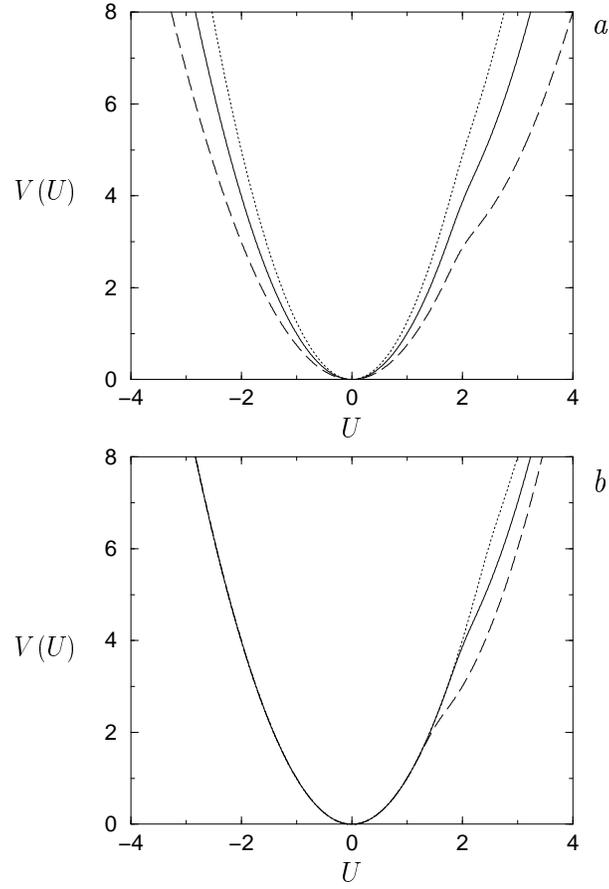}
}
\caption[f]{\label{fig1}
(a) Representation of the potential function $U(V)$ [Eq. (\ref{pf1})] for
$R^{-1}=2$ (continuous line),  $R^{-1}=3$ (dotted line),
and $R^{-1}=1$ (dashed line).
The values of the remaining parameters are 
$C=1$, $\nu=10$, $\theta=2$, $\varepsilon=2$.
(b) Representation of $U(V)$ for $\theta=2$ (continuous line),
$\theta=1.5$  (dotted line),
and $\theta=2.5$ (dashed line).
The values of the remaining parameters are 
$C=1$, $\nu=10$, $R^{-1}=2$, $\varepsilon=2$.
}
\end{figure}

\begin{figure}[th]
\centerline{
\epsfxsize=8.0cm 
\epsffile{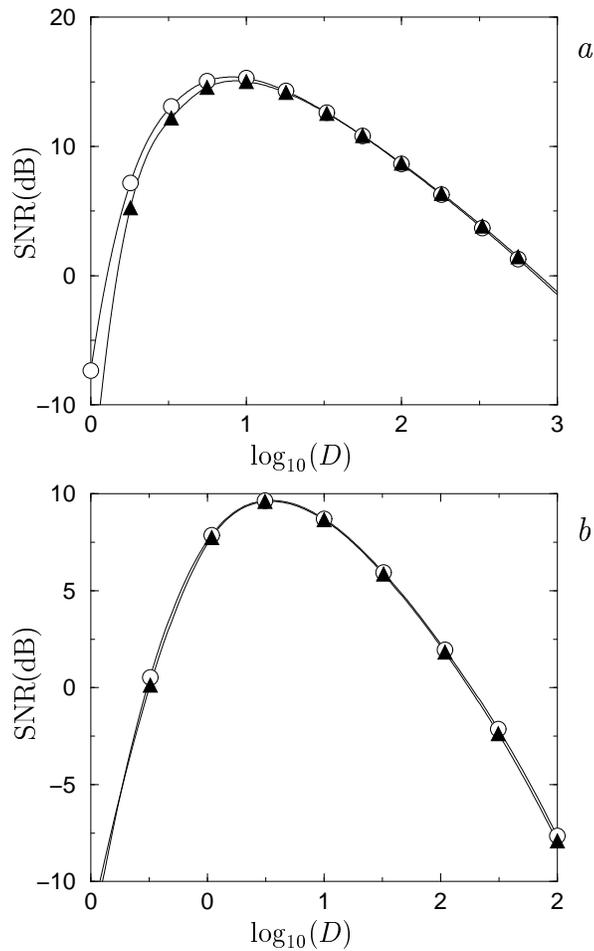}
}
\caption[f]{\label{fig3}
(a) SNR corresponding to Eq. (\ref{lg1}) as a function
of the noise level
for $\omega_0/2\pi=0.1$ (circles) and $\omega_0/2\pi=0.01$ (triangles).
The values of the remaining parameters are $\kappa=2$, $\alpha=0.5$,
$\varepsilon=2$, $x_0=2$, and $\nu=10$. 
(b) SNR corresponding to Eq. (\ref{lg2}) as a function
of the noise level
for $\omega_0/2\pi=0.1$ (circles) and $\omega_0/2\pi=0.01$ (triangles).
The values of the remaining parameters are $\kappa=2$, $\alpha=0.5$,
$\varepsilon=2$, $x_0=2$, and $\nu=10$.
In both cases the solid line is a guide for the eye.
}
\end{figure}

In order to analyze the case in which the oscillations and the nonlinear
term are small, but not infinitesimal, we have numerically integrated the
corresponding equations following a standard second order Runge-Kutta
method for stochastic differential equations \cite{sde1,sde2}.
The Langevin equation we have integrated is the one that corresponds to
the Fokker-Plank equation [Eq. (\ref{FP1})] and is given by
\begin{equation}
\label{lg1}
{dx \over dt}=-k(t)x+\varepsilon [1+e^{-\nu (x-x_0)}]^{-1}+\xi(t) \;\; ,
\end{equation}
where $\xi(t)$ is Gaussian white noise with zero mean, and 
correlation function $\left<\xi(t)\xi(t^\prime)\right>
=2D\delta(t-t^\prime)$.
Here, small means that the effects of the nonlinear term are not
large enough in order for the system to become bistable.
The potential function giving the dynamics of the system is depicted
in Fig. \ref{fig1}a,
for different values of the resistance.
In Fig. \ref{fig3}a we have shown the behavior of the SNR as a function
of the noise level $D$ for two frequencies. The values of
the remaining parameters are the same as the
corresponding ones to the potential function
of Fig. \ref{fig1}a. This figure clearly exhibits a maximum in the SNR.

When the nonlinear term is large enough, the system may display
bistability. One state corresponds to all neurons at rest and
the other to active neurons.

\begin{figure}[th]
\centerline{
\epsfxsize=8.0cm 
\epsffile{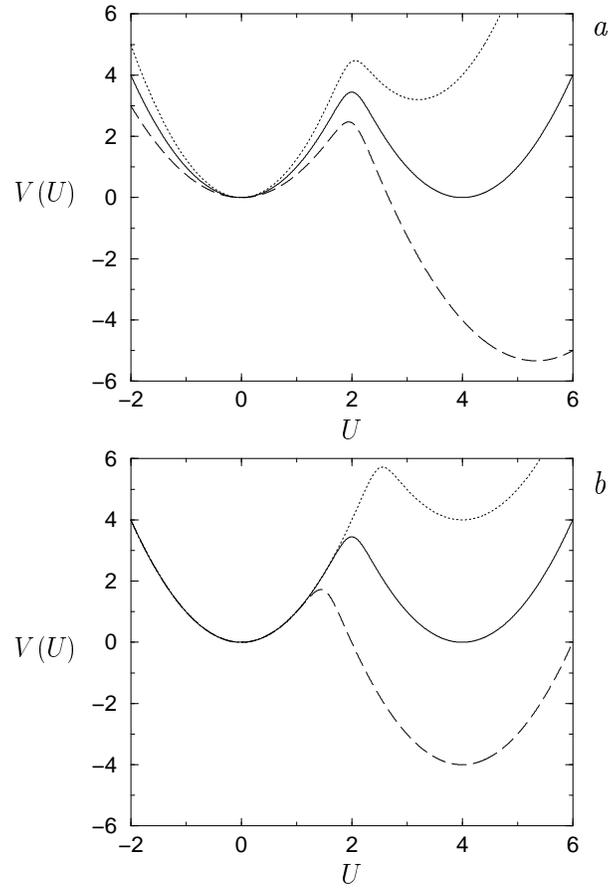}
}
\caption[f]{\label{fig2}
(a) Representation of the potential function $U(V)$ [Eq. (\ref{pf1})].
Same situation as in Fig. \ref{fig1}a, but $\varepsilon=8$.
(b) Representation of $U(V)$.
Same situation as in Fig. \ref{fig1}b, but $\varepsilon=8$.
}
\end{figure}

In Fig. \ref{fig2}a we have displayed the potential function associated
to the Eq. (\ref{lg1}), when the resistance varies for
values of the parameters corresponding
to the bistable situation.
Note that when
the resistance depends on time, the position of the minimum
corresponding to the active state also changes
periodically in time.

\begin{figure}[th]
\centerline{
\epsfxsize=8.0cm 
\epsffile{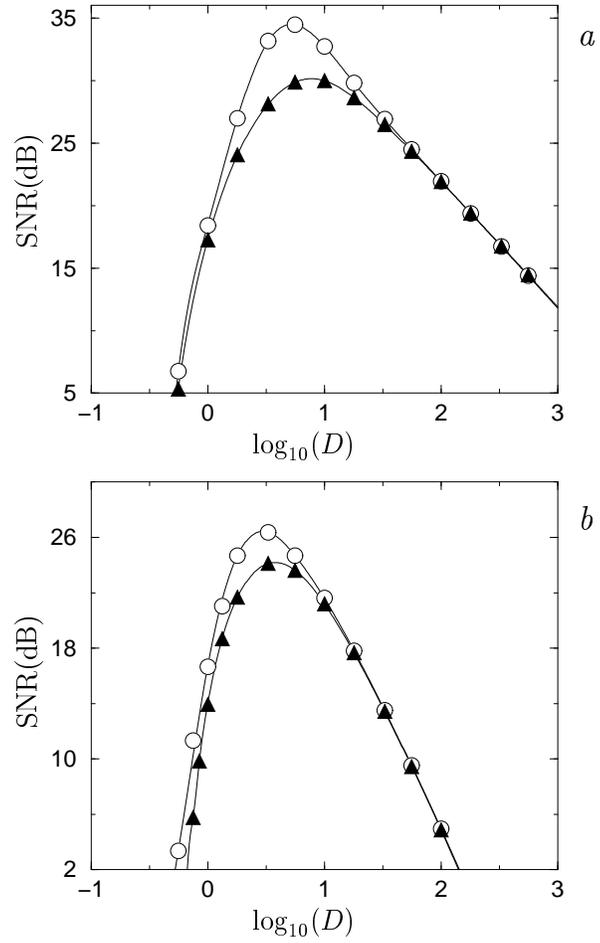}
}
\caption[f]{\label{fig4}
(a) Same situation as in Fig. \ref{fig3}a, but $\varepsilon=8$.
(b) Same situation as in Fig. \ref{fig3}b, but $\varepsilon=8$.
}
\end{figure}

In order to analyze this situation, we have numerically integrated the
corresponding equations as in the previous situation where the 
potential function is monostable.
Fig. \ref{fig4}a displays the SNR which exhibits a maximum at a nonzero
noise level. 

\begin{figure}[th]
\centerline{
\epsfxsize=8.0cm 
\epsffile{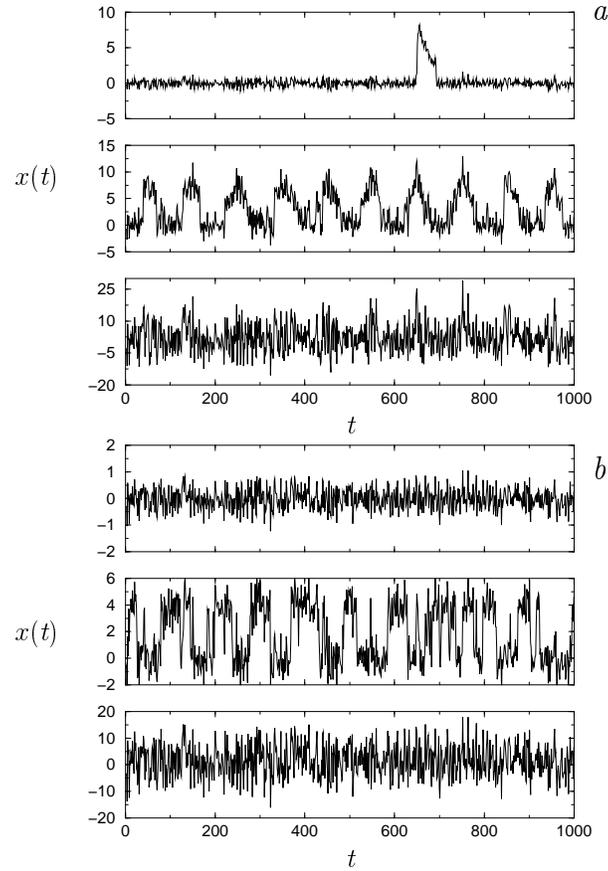}
}
\caption[f]{\label{fig5}
(a) Time series corresponding to Eq. (\ref{lg1})
for $D=0.56$ (top), $D=5.6$ (middle), $D=100$ (bottom).
The values of the remaining parameters are $\omega_0/2\pi=0.01$,
$\kappa=2$, $\alpha=0.5$,
$\varepsilon=8$, $x_0=2$, and $\nu=10$. 
(b) Time series corresponding to Eq. (\ref{lg2})
for $D=0.56$ (top), $D=3.3$ (middle), $D=100$ (bottom).
The values of the remaining parameters are $\omega_0/2\pi=0.01$,
$\kappa=2$, $\alpha=0.5$,
$\varepsilon=8$, $x_0=2$, and $\nu=10$. 
}
\end{figure}

In view to illustrate how the system behaves, in Fig. \ref{fig5}a
we have shown
three time series for different noise levels.
This figure clearly manifests the presence of an optimum noise level,
at which the response of the system is enhanced, and 
the displacement of the minima corresponding to
active neurons.

\section{Variation of the threshold}\label{sec4}

In the previous analysis we have studied the case
in which the resistance of the neuron undergoes oscillations.
Another possibility of temporal variation are oscillations
in the parameter $\theta$.
Explicitly, the dynamics
corresponding to this situation is described again by Eq. (\ref{FP1}), but now
$F_0=-\kappa x$ and $F_1 = [1+e^{-\nu (x-x_0-\alpha\cos(\omega_0 t))}]^{-1}$.
By using the same assumptions about the parameters $\alpha$, $\varepsilon$,
$\omega_0$, and the gain $\nu$ introduced previously, we will now
proceed in a similar fashion.

It is easy to see that
in this case $P_0$ does not depend on time:
\begin{equation}
P_0=\sqrt{\kappa  \over 2\pi D}e^{-\kappa x^2 /2 D}
\;\; ,
\end{equation}
and the correction to the probability density due to the
nonlinear term is given by
\begin{eqnarray}
\nonumber
P_1 &=& {1 \over 2D }{P_0}\{[x-x_0-\alpha\cos(\omega_0 t)]
\Theta(x-x_0-\alpha\cos(\omega_0 t)) \\
\label{P1b}
& & -\int^\infty_{x_0+\alpha\cos(\omega_0 t)}
[x-x_0-\alpha\cos(\omega_0 t)]{P_0(x)}dx\}
\;\; .
\end{eqnarray}
The averaged value of $x$ is then
\begin{equation}
\left<x\right>_1=\int_{x_0+\alpha\cos(\omega_0 t)}^\infty
x{1 \over 2D }{P_0(x)}[x-x_0-\alpha\cos(\omega_0 t)]dx
\;\; .
\end{equation}
Note that for symmetry reasons, the integral in
Eq. (\ref{P1b}) gives a null
contribution to $\left<x\right>_1$.
By expanding in the parameter $\alpha$ around its zero value,
we then obtain Eq. (\ref{sus}).
In this case the susceptibility reads
\begin{equation}
{\cal A} 
= -\int_{x_0}^\infty x{1 \over 2D }{P_0(x)}dx
=-{1 \over \sqrt{ 8\pi \kappa D}}e^{-\kappa x_0^2 /2 D}
\;\; .
\end{equation}

The noise term is the same as in the previous case, then we obtain
the following expressions for the signal, noise and SNR
\begin{eqnarray}
S(\omega_0) &=& {\pi \over 2}(\alpha\epsilon{\cal A})^2
={1 \over 16\pi \kappa D}(\alpha\epsilon)^2
e^{-\kappa x_0^2 /D} \;\; , \\
%
N(\omega_0) &= & { D \over 2\kappa ^2} \;\; , \\
\mbox{SNR} &=& \pi(\alpha\epsilon{\cal A})^2 {\kappa ^2\over D} 
={\kappa (\alpha\epsilon)^2 \over 8\pi D^2}
e^{-\kappa x_0^2 /D}
\;\; .
\end{eqnarray}
This last expression clearly shows that the SNR has a maximum
at a non-zero noise level then making the presence of SR manifest.

As we did in the previous situation, 
to study the case in which the oscillations and the nonlinear
term are not infinitesimal, we have numerically solved
the corresponding Langevin equation
by using the procedure outlined above.
In this case it reads
\begin{equation}
\label{lg2}
{dx \over dt}=-\kappa x+
\varepsilon[1+e^{-\nu (x-x_0-\alpha\cos(\omega_0 t))}]^{-1}+\xi(t) \;\; ,
\end{equation}
where $\xi(t)$ is the same noise as the one defined previously.
The explicit situation we have considered is given through
the potential function displayed in Fig. \ref{fig1}b.
In  Fig. \ref{fig3}b we have shown the behavior of the SNR as a function
of the noise level $D$ for two frequencies. The values of
the remaining parameters are the same as those
corresponding to the potential function
of Fig. \ref{fig1}b.
In this case the SNR also exhibits a maximum for the two frequencies.

Bistability may also be present in this case. In Fig. \ref{fig2}b
we have represented the potential function when periodic modulations act
through the threshold, for values of the parameters
corresponding to the bistable situation. In contrast with the
case of oscillations of the resistance, in which the
minimum corresponding to the active state varies its position,
when the threshold oscillates, the two minima always remain at
the same transmenbrane potential. 

This situation can also be analyzed through numerical integration.
Fig. \ref{fig4}b displays the SNR for the periodic modulation
we are considering.
This quantity exhibits a maximum for the two frequencies.
Finally, in Fig \ref{fig5}b we have also shown three time series for different
noise levels. It is worth emphasizing that, for noise levels
close to the optimum value, the time series
look as those corresponding to the usual bistable
quartic potential \cite{Ma1,Ma2}.

\section{Discussion}\label{sec5}

In this paper we have analyzed
how noise affects the behavior of a neural medium when 
it is periodically
modulated. We have found that the occurrence of noise
may play a constructive role since an optimized amount of it
may contribute to enhancing the response of the system.
Under some circumstances, the presence of noise is responsible
for the appearance of oscillations which otherwise would not
be manifested.
In this regard, the analysis of macroscopic neural dynamics
obtained from electroencephalograms has
been a matter of debate over the last decade \cite{Bas}.
It is accepted that the
activity of the brain cortex shows low-dimensional traits, though the
exact nature of the phenomenon itself is far from being clear.
Here we have shown that noise, sometimes not considered, could
give rise to a coherent behavior of the system,
then playing an important role in neural dynamics.

The mean field model we have proposed,
although constituting an oversimplified picture of a
thalamo-cortical network, might be
a first step in our understanding of how noise and nonlinearities can
generate interesting macroscopic outcomes.
Further developments must include spatial effects as well as the
consideration of activatory and inhibitory populations of neurons.

On what concerns to the phenomenon of SR itself,
an important aspect that should be emphasized is the fact that
the model we have presented
exhibits SR in both monostable and bistable situations,
depending on the values of the parameters.
This remarkable feature contrasts with previous studies for
which SR has been found only for monostable or bistable systems.
Moreover, some bistable systems undergoing SR may become
monostable, but under this circumstance SR does not take place.
In fact, the model under consideration describes
a system with
a threshold, accounting for the firing of the neurons
that may behave as a monostable or a bistable system.
Therefore, we have envisaged a model that
may exhibit three different situations
(monostable, bistable, and threshold)
where the phenomenon occurs.  

\section*{Acknowledgments}

This work was
supported by DGICYT of the Spanish Government under Grants Nos.
PB95-0881 and PB94-1195.
J. M. G. V. wishes to thank Generalitat
de Catalunya for financial support.

\end{document}